\documentclass{aa}

\usepackage{txfonts}
\usepackage{multirow}
\usepackage{hyperref}
\usepackage{xcolor}

\definecolor{myblue}{HTML}{387EE1}
\definecolor{mygreen}{HTML}{00964B}
\definecolor{myred}{HTML}{D9321F}

\hypersetup{
	colorlinks=true,
	linkcolor=myred,
	urlcolor=myblue,
	citecolor=mygreen
}

\newcommand{\quotes}[1]{``#1''}

\begin{document}

\title{Catalog of quasars from the Kilo-Degree Survey Data Release 3}

\author{
	S.~Nakoneczny\inst{\ref{ncbj}} \and
	M.~Bilicki\inst{\ref{leiden},\ref{cft}} \and
	A.~Solarz\inst{\ref{ncbj}} \and
	A.~Pollo\inst{\ref{ncbj},\ref{uj}} \and
	N.~Maddox\inst{\ref{lmu}}  \and
	C.~Spiniello\inst{\ref{inaf}} \and
	M.~Brescia\inst{\ref{inaf}}	\and
	N.R.~Napolitano\inst{\ref{sun},\ref{inaf}}
}

\institute{
	National Centre for Nuclear Research, Astrophysics Division, ul.~Ho\.{z}a 69, 00-681 Warsaw, Poland \label{ncbj} \and
	Leiden Observatory, Leiden University, P.O. Box 9513, NL-2300 RA Leiden, The Netherlands \label{leiden} \and
	Center for Theoretical Physics, Polish Academy of Sciences, al. Lotnik\'{o}w 32/46, 02-668, Warsaw, Poland \label{cft} \and 
	Astronomical Observatory of the Jagiellonian University, 31-007 Krak\'{o}w, Poland  \label{uj} \and
	Faculty of Physics, Ludwig-Maximilians-Universit\"at, Scheinerstr. 1, 81679 Munich, Germany \label{lmu} \and
	INAF -- Osservatorio Astronomico di Capodimonte, Salita Moiariello, 16, I-80131 Napoli, Italy \label{inaf} \and
	School of Physics and Astronomy, Sun Yat-sen University, Guangzhou 519082, Zhuhai Campus, P.R. China \label{sun}
} 

\authorrunning{S.~Nakoneczny et al.}

\offprints{S.~Nakoneczny, \email{\url{szymon.nakoneczny@ncbj.gov.pl}}.}

\abstract{We present a catalog of quasars selected from broad-band photometric $ugri$ data of the Kilo-Degree Survey Data Release 3 (KiDS DR3). The QSOs are identified by the random forest (RF) supervised machine learning model, trained on Sloan Digital Sky Survey (SDSS) DR14 spectroscopic data. We first cleaned the input KiDS data of entries with excessively noisy, missing or otherwise problematic measurements. Applying a feature importance analysis, we then tune the algorithm and identify in the KiDS multiband catalog the 17 most useful features for the classification, namely magnitudes, colors, magnitude ratios, and the stellarity index. We used the t-SNE algorithm to map the multidimensional photometric data onto 2D planes and compare the coverage of the training and inference sets. We limited the inference set to $r<22$ to avoid extrapolation beyond the feature space covered by training, as the SDSS spectroscopic sample is considerably shallower than KiDS. This gives 3.4 million objects in the final inference sample, from which the random forest identified 190,000 quasar candidates. Accuracy of 97\% (percentage of correctly classified objects), purity of 91\% (percentage of true quasars within the objects classified as such), and completeness of 87\% (detection ratio of all true quasars), as derived from a test set extracted from SDSS and not used in the training, are confirmed by comparison with external spectroscopic and photometric QSO catalogs overlapping with the KiDS footprint. The robustness of our results is strengthened by number counts of the quasar candidates in the $r$ band, as well as by their mid-infrared colors available from the Wide-field Infrared Survey Explorer (WISE). An analysis of parallaxes and proper motions of our QSO candidates found also in Gaia DR2 suggests that a probability cut of $p_\mathrm{QSO} > 0.8$ is optimal for purity, whereas $p_\mathrm{QSO} > 0.7$ is preferable for better completeness. Our study presents the first comprehensive quasar selection from deep high-quality KiDS data and will serve as the basis for versatile studies of the QSO population detected by this survey. We publicly release the resulting catalog at \url{http://kids.strw.leidenuniv.nl/DR3/quasarcatalog.php}, and the code at \url{https://github.com/snakoneczny/kids-quasars}.}


\maketitle

\section{Introduction}

One of the key goals of ongoing and planned wide-angle sky surveys is to map the large-scale structure (LSS) of the Universe and derive various cosmological constraints, using different probes such as galaxy clustering or gravitational lensing. The building blocks of the LSS are galaxies, and among them those with active galactic nuclei (AGN) stand out. Presence of an AGN is a signature of a growing supermassive black hole (SMBH) at the center of the galaxy \citep[e.g.,][]{Kormendy:2013}. AGN luminosity, contrary to its host galaxy, does not come from stellar radiation but rather from energy released during the accretion of the material onto the SMBH. Various accompanying phenomena, including jets launched orthogonally to the accretion disk, make the AGNs very luminous. This means that they can be detected from large, cosmological distances, and as they often outshine their host galaxies, they are observed as point-like quasars\footnote{In this paper we use the terms \quotes{quasar} and \quotes{QSO} interchangeably, to denote any galaxy which has a bright, actively accreting nucleus.}.

Quasars  typically reside in very massive dark matter haloes of masses above $10^{12} M_{\odot}$ \citep[e.g.,][]{Eftekharzadeh:2015,DiPompeo:2016}, and trace the peaks of the underlying matter field. In cosmological terms this means that QSOs are highly biased tracers of the LSS \citep[e.g.,][]{DiPompeo:2014,Laurent:2017}. Together with their large intrinsic luminosities, this makes QSOs very useful cosmological probes: they can be detected up to very high redshifts and their measured clustering amplitude is considerably higher than that of field galaxies. On the other hand, at any cosmic epoch quasars are much more sparsely distributed than inactive galaxies, and are therefore sufficiently large and cosmologically useful samples of QSOs require wide-angle surveys covering large volumes of the Universe. And indeed, only as a result of dedicated programs such as the 2dF QSO Redshift Survey \citep[2QZ,][]{Croom:2004} or the Sloan Digital Sky Survey \citep[SDSS,][]{York:2000} do we now have catalogs of spectroscopically confirmed QSOs counting $\sim 10^4$-$10^5$ objects.

The most robust way of identifying quasars is via their spectra, which present specific features such as broad emission lines and strong signatures of emission line ratios like [OIII]$\lambda 5007$/H$\beta$, [NII]$\lambda$6584/H$\alpha$ \citep[e.g.,][]{Kauffmann:2003,Kewley:2013}.
This helps to easily distinguish them from inactive galaxies as well as from other point-like sources, namely Galactic stars. This is the approach taken by dedicated spectroscopic QSO surveys such as the 2QZ, the 2dF-SDSS LRG and QSO survey \citep[2SLAQ,][]{Croom:2009} or the SDSS \citep[e.g.,][]{Paris:2018}, as well as the forthcoming DESI \citep{DESI:2016} and 4MOST \citep{deJong:2011}. However, the spectroscopic quasar confirmation method has its limitations as it is very expensive in terms of telescope time, requiring long integrations due to typically low observed fluxes of such sources (related to their enormous distances, despite their large intrinsic luminosities). For this reason methods are being developed to identify quasars from datasets that are more readily available, that is wide-angle broad-band photometric samples. 

The imaging-based QSO selection approaches range from applying color cuts to photometric data \citep[e.g.,][]{Warren:2000,Maddox:2008,Edelson:2012,Stern:2012,Wu:2012,Secrest:2015,Assef:2018} through more sophisticated probabilistic methods \citep[][etc.]{Richards:2004,Richards:2009a,Richards:2009b,Bovy:2011,Bovy:2012,DiPompeo:2015,Richards:2015} to machine learning (ML) based automated classification approaches \citep[e.g.,][]{Brescia:2015,Carrasco:2015,Kurcz:2016}. In general, these approaches take advantage of the fact that quasars, or more generally AGNs, display colors at various wavelengths which are distinct from those of inactive galaxies as well as of stars. However, the reliability of QSO detection depends on the available number of colors and magnitudes. It is therefore desirable to work in multidimensional parameter spaces where quasar selection becomes more efficient, for example by combining optical and infrared (IR) information. In such a situation, however, traditional color division becomes challenging due to difficulties with projecting $N$ dimensions ($N$D) onto 2- or at most 3D. For those reasons, automated QSO detection via ML is gaining on popularity in the recent years.

Wide-angle quasar samples, especially if they include some distance information such as from spectroscopic or photometric redshifts, have numerous applications. Photometrically selected QSOs are especially useful for studies where high number density and completeness, not readily available from spectroscopic samples, are crucial. These applications include tomographic angular clustering \citep[e.g.,][]{Leistedt:2014,Ho:2015}, analyses of cosmic magnification \citep[e.g.,][]{Scranton:2005}, measurements of halo masses \citep[e.g.,][]{DiPompeo:2017}, cross-correlations with various cosmological backgrounds 
\citep[e.g.,][and references therein]{Sherwin:2012,Cuoco:2017,Stolzner:2018}, and even calibrating the reference frames for Galactic studies \citep[e.g.,][]{Lindegren:2018}.

In this work we explore quasar detection in one of the deepest ongoing wide-angle photometric surveys, the Kilo-Degree Survey\footnote{\url{ http://kids.strw.leidenuniv.nl/}} \citep[KiDS,][]{deJong:2013}. 
The depth of KiDS, $\sim25$ mag in the $r$ band ($5\sigma$), its multiwavelength $ugri$ coverage, and availability of overlapping VIKING data at a similar depth \citep{Edge:2013}, make this survey an ideal resource for quasar science. This remains however a very much uncharted territory in KiDS, and only three studies so far presented QSO-related analyses based on this survey: \cite{Venemans:2015} focused on very high-redshift ($z\sim6$) quasars found in a combination of KiDS and VIKING data, \cite{Heintz:2018} studied a heavily reddened QSO identified in KiDS+VIKING,  while \cite{Spiniello:2018} selected QSO-like objects over the KIDS DR3 footprint to search for strong-lensing systems.

Here we make the first step towards systematic studies of the KiDS quasar population by presenting automated detection of QSOs in the most recent KiDS public Data Release 3 \citep[DR3,][]{deJong:2017}. For that purpose we employ one of the most widely used supervised machine learning algorithms, random forest, to detect QSOs in KiDS imaging in an automated way. The model is trained and validated on spectroscopic quasar samples which overlap with the KiDS DR3 footprint. We put special emphasis on selecting the most informative features for the classification task, as well as on appropriate trimming of the target dataset to match the training feature space and avoid unreliable extrapolation. This is also aided by analysis of two-dimensional projection of the high-dimensional feature space. The trained algorithm is then applied on the photometric KiDS data, and the robustness of the resulting quasar selection is verified against various external catalogs: point sources from the Gaia survey, as well as spectroscopic and photometric quasar catalogs, which were not used for training or validation. We also verify the number counts as well as mid-IR colors of the final QSO catalog.

The paper is organized as follows. In \S\ref{section_data} we describe the data used for classification, and how it was prepared to construct the inference sample (from KiDS) as well as the training set (based on SDSS). Section \ref{section_classification_pipeline} provides details of the classification pipeline, including the random forest machine-learning model, performance evaluation methodology, model tuning, feature selection and feature space limitation to match the training data. In \S\ref{section_evaluation} we discuss the results of quasar classification in KiDS DR3, and quantify its performance both by internal tests with SDSS data, as well as by comparing the output with external star and quasar datasets. In the final \S\ref{section_conclusions} we conclude and mention future prospects for KiDS quasar selection.

\section{Data}
\label{section_data}

In this Section we describe the data used for the quasar selection and for the subsequent validation. We aim to select QSO candidates from the photometric data derived from the KiDS DR3. As our classification method learns in a supervised manner, it requires a training set with known object labels which we obtain from SDSS DR14 spectroscopic data cross-matched with KiDS.

\subsection{Inference set from the Kilo-Degree Survey}
	
KiDS is an ongoing  wide-field imaging survey using four optical broad-band filters, $ugri$, employing the 268 Megapixel OmegaCAM camera \citep{Kuijken:2011} at the VLT Survey Telescope \citep[VST][]{Capaccioli:2012}. KiDS provides data of excellent photometric quality, with $5\sigma$ depth of $r\sim25$ mag and typical seeing of $\sim0.7"$ in the $r$ band. In this work we make use of its most recent public Data Release 3 \citep{deJong:2017} which covers $\sim 447$ deg$^2$ and includes almost 49 million sources in its multiband catalog that we use as the parent dataset.

\subsubsection{Basic features used in the classification}

The main features used in the classification process come directly from the KiDS catalog and consist of the \textit{ugri} magnitudes and corresponding colors. As detailed in \cite{deJong:2017}, KiDS data processing provides various photometric measurements of detected sources. For our purpose we need robust measurements of point source photometry, we therefore use the GAaP magnitudes (Gaussian Aperture and PSF, \citealt{Kuijken:2008}) which are designed to compensate for seeing variations among different filters. Together with additional photometric homogenization across the survey area, these measurements provide precisely calibrated fluxes and colors \citep{Kuijken:2015}. Combining the four magnitudes and six colors (one for every magnitude pair) results in ten basic features. Although using both magnitudes and colors seems redundant with respect to using for example only magnitudes or only colors, such redundancy does improve classification results. As detailed later in \S\ref{subsection_feature_sel}, we also tested ratios of magnitudes as additional parameters for the classification. Together with the stellarity index \texttt{CLASS\_STAR} (see below), the magnitudes, colors and magnitude ratios constitute the eventual 17D feature space for classification, in which only the most relevant features are used.

We note that although a significant fraction of KiDS sources are stars, we always use magnitudes and colors corrected for Galactic extinction, as our focus is to detect extragalactic sources. The measurements are also corrected for the \quotes{zero-point offset} to ensure flux uniformity over the entire DR3 area\footnote{These offsets are on average $\sim0.03$ mag in the $u$ and $i$ bands, and $<0.01$ mag in the $g$ and $r$ bands. See \cite{deJong:2017} for details.}.

Another parameter used in the classification process, which turns out to be very important for the performance (\S\ref{subsection_feature_sel}), is \texttt{CLASS\_STAR}. This is a continuous stellarity index derived within the KiDS data processing pipeline using SExtractor \citep{Bertin:1996}, which describes the degree to which a source is extended. It takes values between 1 (point-like, a star) and 0 (extended, typically a galaxy). Most quasars, except for the rare ones with a clearly visible extended host, are point-like and therefore have high values of \texttt{CLASS\_STAR}. The reliability of this parameter for separation of point sources from extended ones depends on the signal-to-noise (S/N) of the objects and resolution of the imaging, therefore it may fail to identify faint and small galaxies (in terms of apparent values). However, the data used for the quasar detection in this work are limited to a relatively bright and high S/N subsample of KiDS DR3, where \texttt{CLASS\_STAR} is robust in separating these two main source classes. Indeed, we have verified that for the training set  with known labels (\S\ref{subsection_training_set})\footnote{This applies equally to validation and test sets, as they are chosen randomly from the general SDSS labeled sample.}, separation at \texttt{CLASS\_STAR} of 0.5 would leave a negligible fraction ($0.5\%$) of galaxies marked as \quotes{point-like} and a small number ($1.9\%$) of stars identified as \quotes{extended}. As far as the training-set QSOs are concerned, the vast majority of them have $\mathtt{CLASS\_STAR}>0.6$. Some quasars with resolved hosts may  still be detected as extended, especially in a survey with such excellent angular resolution as KiDS. We therefore decided to test the usefulness of \texttt{CLASS\_STAR} in the quasar automated selection and indeed found it very helpful. This is discussed in more detail in \S\ref{subsection_feature_sel}. We would like to emphasize, however, that no a priori cut on \texttt{CLASS\_STAR} is made in either the training or inference sample. The QSO classification algorithm filters out most of the extended sources, but does identify a fraction of them as quasars. Roughly 1\% of the objects with $\mathtt{CLASS\_STAR}<0.5$ are assigned a QSO label (20k out of 2M). This means that $\sim$11\% of all our quasar candidates were classified by KiDS as extended.

The KiDS data processing pipeline provides also another star/galaxy separator, \texttt{SG2DPHOT} \citep{deJong:2015}, which uses the $r$-band source morphology and generally is more robust than \texttt{CLASS\_STAR}. We have however found that in our particular application, using \texttt{SG2DPHOT} instead of \texttt{CLASS\_STAR} gives slightly worse classification results. Part of the reason might be that \texttt{SG2DPHOT} is a discrete parameter so it provides less information to the model than continuous \texttt{CLASS\_STAR}.

An additional potentially useful feature in the classification process could be photometric redshifts (photo-$z$s). However, KiDS DR3 photo-$z$s were optimized for galaxies \citep{deJong:2017,Bilicki:2018} and are unreliable for quasars, as we indeed verified for overlapping spectroscopic QSOs. In future work we will address the issue of deriving more robust QSO photo-$z$s in KiDS as well as estimating them jointly with object classification \citep[see e.g.,][]{Yeche:2010}.

\subsubsection{Data preprocessing}
\label{subsub_data_preproc}

Machine-learning methods, such as the one employed in this paper, require data of adequate quality in order to perform reliably. In addition, in supervised learning, it is desirable to avoid extrapolation beyond the feature space covered by the available training set. For those reasons, we apply appropriate cleaning and cuts on KiDS DR3 data as specified below, to ensure reliable quasar classification.

\begin{table}
	\caption{Numbers of objects left in the KiDS DR3 inference data after the subsequent preprocessing steps. See text for details of these cuts.}
	\centering
	\begin{tabular}{c c c}
		\hline\hline
		& objects left & \% of all data \\
		\hline
		all KiDS DR3 sources & 49M & 100\% \\
		keep only four-band detections & 40M & 80\% \\
		cut at limiting magnitudes & \multirow{2}{*}{11M} & \multirow{2}{*}{22\%} \\
		\& remove errors > 1 mag & & \\
		clean up image flags & 9M & 18\% \\
		cut at $r < 22$ & 3.4M & 6.8\% \\
		\hline
	\end{tabular}
	\label{table:data_kids}
\end{table}	

\begin{figure*}
	\resizebox{\hsize}{!}{\includegraphics{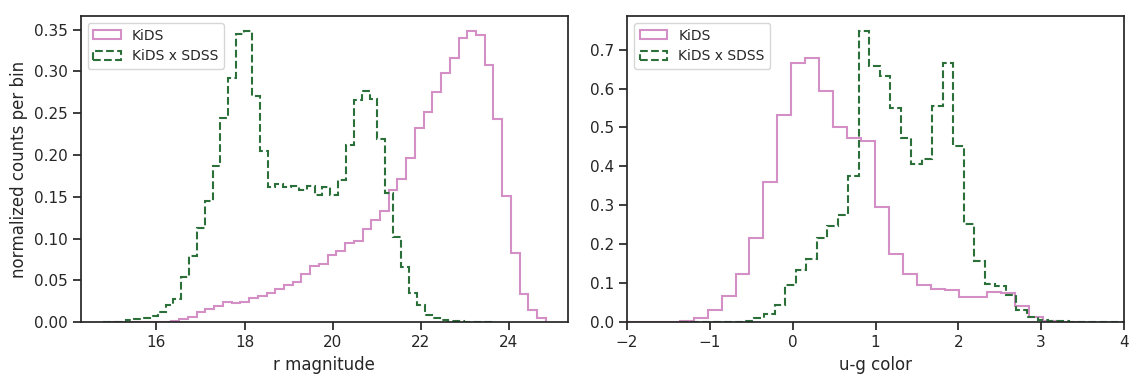}}
	\caption{Normalized histograms of the $r$-band magnitude (left) and the $u-g$ color (right) for the KiDS inference dataset (pink solid) and the KiDS $\times$ SDSS training set (green dashed), both before applying the $r<22$ mag cut. The bimodality of the matched sample is related to SDSS spectroscopic target preselections, preserved by the cross-match with the much deeper KiDS.}
	\label{fig:r_feature_space}
\end{figure*}

\begin{enumerate}
	
\item \textit{Keep only four-band detections}

A fraction of KiDS DR3 sources do not have all the four bands measured. As Machine-learning models require all employed features to bear a numerical representation, using sources with missing features would require assigning some artificial values to relevant magnitudes and colors. As already one missing magnitude means three colors are not available, even in such a minimal scenario it would significantly reduce the available information. As we show below, colors are in fact among the most important features in our classification procedure, and we cannot afford to lose them. Therefore, we only use objects with all the four $ugri$ bands measured. This step removes about 20\% of the catalog entries, leaving roughly 40 million sources (Table \ref{table:data_kids}).\\
	
\item \textit{Ensure sufficient signal-to-noise ratio}

To avoid working with excessively noisy data which could significantly affect classification performance, we first cut the KiDS data at the nominal limiting magnitude levels, which are 24.3, 25.1, 24.9, 23.8 in \textit{ugri}, respectively. Additionally, we require the photometric errors in each band to be smaller than 1 mag, which roughly corresponds to S/N of 1. These two cuts applied simultaneously on all the bands (i.e., as a joint condition) are the most strict among other cleaning requirements and together with the above item \#1, leave $\sim11$ million KiDS DR3 objects. However, we note that due to training set limitations, a further and stricter cut on the $r$ band magnitude is applied below.\\

\item \textit{Remove flagged objects}
	
The KiDS data processing pipeline provides various flags for detected objects, indicating possible issues with photometry (see \citealt{deJong:2017} for details). In this work, we take into account \texttt{IMAFLAGS\_ISO\_band} which are flags delivered by the KiDS pipeline and include information about critical areas in the images, which likely corrupt single source photometry. Issues such as star halos and spikes are automatically detected using an algorithm that first finds star positions from the saturation map, and then builds models of star halos and spikes taking into account the telescope orientation and the position in the focal plane (the Pullecenella mask procedure, \citealt{deJong:2015}). We have carefully examined several images of objects with these flags set and decided to remove sources indicated by any of the bits\footnote{These are: 1 - saturation spike, 2 - saturation core, 4 - diffraction spike, 8 - primary reflection halo, 16 - secondary reflection halo, 32 - tertiary reflection halo, 64 - bad pixel.}  except for the one for manual masking, which was valid only for KiDS DR1 and DR2 \citep{deJong:2015}. We have also considered \texttt{FLAG\_band} (a SExtractor flag indicating possible issues in the extraction of the object) but found no significant deterioration in classification quality after including sources marked by this flag in the training and inference process. This cleanup step removes a non-negligible number of sources, about 2 million out of 11 million that were left after the previous step \#2.\\	

\item \textit{Trim target data to match the training set}

The previous steps of data cleaning were of a general nature to ensure adequate KiDS data quality for classification purposes. A final step is however required and it is specific to the training sample used. Namely, the SDSS DR14 spectroscopic  training set does not reach beyond $r\sim22$ mag (see Fig.~\ref{fig:r_feature_space} left panel). Using significantly deeper inference data than for the training set would require extrapolation, which for supervised ML might be unreliable and more importantly, its performance would be difficult to evaluate. Therefore, the target KiDS data must be trimmed to limit the feature space to ranges covered by training. As the default cut we adopt $r<22$, although we have also experimented  with a more permissive cut in the $u-g$ color, as matching the training and inference sets in this parameter leads to the removal of fewer objects (see Fig.~\ref{fig:r_feature_space} right panel). We provide more discussion on feature space limitation and our final choices in \S\ref{subsection_feature_space}.

This particular cut significantly limits the size of usable data for the classification, leaving us with 3.4 million KiDS DR3 objects. Such a cut would in fact make unnecessary the condition on the sufficient S/N described in item \#2 above, as KiDS sources with $r<22$ mag typically have very high S/N in all the bands. However, we keep this condition separately, as it is related to the characteristics of the specific training set used.

We note that with future deeper QSO training sets, such as from the final SDSS-IV \citep{Blanton:2017} or DESI \citep{DESI:2016}, it will be possible to extend the range of the usable feature space and therefore increase the number density of robustly identified quasars in KiDS. 

\end{enumerate}

Table \ref{table:data_kids} summarizes all the preprocessing steps which led to the creation of the final dataset on which our quasar catalog is based. We denote this dataset as the target or inference sample.

\subsection{Training set from the Sloan Digital Sky Survey}
\label{subsection_training_set}
	
To learn object classification based on photometric data, our machine learning model needs ground-truth labels. In this work, the labels are taken from the SDSS, and in particular from the spectroscopic catalog of its Data Release 14 \citep{Abolfathi:2018}. It includes over 4.8 million sources with one of three labels assigned: star, galaxy, or quasar. To ensure the highest data quality and model performance, from that sample we use only sources with secure redshifts (velocities) by demanding the zWarning parameter to be null. The cross-match between the SDSS and KiDS inference dataset was done with the TOPCAT tool \citep{Taylor:2005} within a matching radius of $1"$.  The final training dataset consists of 12,144 stars, 7,061 quasars and 32,547 galaxies, which totals to 51,752 objects. These relatively low numbers are the consequence of the small sky overlap between SDSS and KiDS DR3, as the common area covers only the KiDS equatorial fields at $ -3\degr < \delta < 3\degr$. In the test phase, from these labeled data we randomly extract the actual training, validation, and test sets, as detailed in \S\ref{subsection_performance_evaluation}. For training the final classification model, the entire spectroscopic cross-matched sample is used. Therefore, whenever parameter distributions or feature space properties are discussed for \quotes{training}, this applies equally to the general training data, as well as to the validation and test sets.

In Fig.~\ref{fig:r_feature_space} we present normalized distributions of the $r$-band magnitudes (left panel) and the $u-g$ color (right panel) for the KiDS inference dataset (pink solid) and the KiDS $\times$ SDSS training set (green dashed), both before applying the $r<22$ mag cut. This shows clearly that the current SDSS data do not probe KiDS beyond the adopted $r$-band cut. On the other hand, the color space is better covered between the inference and training sample, although here we only showed one particular color as an example. The matching between the training and inference (target) set in the multidimensional feature space is discussed in more detail in \S\ref{subsection_feature_space}. The bimodality of the KiDS $\times$ SDSS seen in both histograms is related to preselections of the SDSS spectroscopic targets at the various stages of the survey. In particular the flux-limited ($r<17.77$) complete \quotes{SDSS Main} sample \citep{Strauss:2002} gives the first peak in the $r$-band histogram, while subsequent BOSS color selections used fainter magnitudes \citep{Dawson:2013}. As KiDS is much deeper than any of the SDSS spectroscopic subsamples, the cross-match preserves these properties.

\section{Classification pipeline}
\label{section_classification_pipeline}

In this section, we present random forest which is our choice of algorithm used for quasar classification in KiDS DR3, and provide its most important details. We also describe how its performance is evaluated and the procedure of feature selection. Finally, we analyze the coverage of the feature space by the training and inference data using an advanced visualization technique called t-distributed stochastic neighbor embedding \citep[t-SNE,][]{Maaten:2008}.

We start by defining the ML problem. The goal of this work is to detect quasars in KiDS DR3 data, however the training sample from SDSS provides labels for 3 types of sources: stars, galaxies, and QSOs. These objects usually populate different regions of the feature space which we use, and we have verified that the QSO identification is more robust if the model is formulated as a three-class rather than a binary (QSOs vs. the rest) problem. This is an expected result as in the three-class case we provide the model with more information.

\begin{table}
	\caption{A comparison of the test results for different models, achieved on the SDSS-based test set separated from the training and validation data. See \S\ref{subsection_performance_evaluation} for details of the metrics.}
	\centering
	\begin{tabular}{c c c}
		\hline\hline
		& three-class: accuracy & QSO vs. rest: F1 \\
		\hline
		RF & 96.56\% & 88.67\% \\
		XGB & 96.44\% & 88.12\% \\
		ANN & 96.28\% & 87.63\% \\
		\hline
	\end{tabular}
	\label{table:models}
\end{table}

Several of the most popular classification algorithms were tested, in particular random forest \citep[RF,][]{Breiman:2001}, artificial neural network \citep[ANN,][]{Haykin:1998} and Xtreme Gradient Boosting \citep[XGB,][]{Chen:2016}. The whole pipeline was implemented using the Python language, while model implementations were taken from several sources: RF from the scikit-learn library\footnote{https://scikit-learn.org} \citep{scikit-learn}, ANNs from the Keras library\footnote{https://keras.io} \citep{Chollet:2015} with the Tensorflow backend\footnote{https://www.tensorflow.org} \citep{Tensorflow:2015}, while XGB has a standalone package. In the case of RF, the best results are usually achieved by building fully extended trees with leafs belonging only to one class, which also provided the best results for our work. We chose entropy as the function to measure the quality of a split, and 400 trees in the model as we did not observe any performance gain above this value. For XGB, we obtained good results when using 200 estimators of depth 7 and trained with a 0.1 learning rate, while artificial neural network was built with 2 hidden layers of 20 neurons each, using the rectified linear unit (ReLU) activation function. Table \ref{table:models} shows a comparison between the model performances; for details of the model testing procedure and evaluation metrics see \S\ref{subsection_performance_evaluation}. We observed small differences between the performance of different models, with RF generally performing best. Such model hierarchy and small differences in scores are expected for this kind of a dataset with a rather low number of features and classes to predict. 

We decided to choose random forest as the final classifier. This decision was based not only on the model performance, but also because RF does not require time-consuming selection of the best training parameters. It also provides a measure of feature importance, offering a relatively fast and straightforward way of choosing the most appropriate features. We discuss this in more detail in \S\ref{subsection_feature_sel}.

\subsection{Random forest}
\label{subsection_random_forest}

The random forest \citep[RF,][]{Breiman:2001} is a widely used classification model, with numerous applications in astronomy \citep[e.g][]{ Masci:2014,Hernitschek:2016,Moller:2016}. It belongs to the family of ensemble methods, in which the final model output is based on many decisions of standalone models. The basic idea is that a decision made together by many voters, which individually can significantly differ, is more accurate than from a single vote. This can be related to many real-life situations, where important decisions are made after consulting with many specialists from different fields. Such an approach can in particular prevent us from making an incorrect decision based on the knowledge of just one specialist, which in ML relates to the problem of overfitting. In the case of RF, the basic single model is the decision tree.

\subsubsection{Decision trees}

The decision tree (DT) works by sequentially dividing a dataset into subsets which maximize homogeneity with respect to ground-truth labels. The best option, implemented in the libraries, is a binary DT where two subsets are created at each split. Assuming a set with observations belonging to different classes, we calculate the probability of class $i$ as
\begin{equation}
p_i = \frac{n_i}{N}\;,
\end{equation}
where $n_i$ is the number of data points belonging to class $i$ and $N$ is the total number of points. Dataset impurity can then be quantified using measures such as entropy, equal to $- \sum_{i} p_i \log_2 p_i$, and Gini index, expressed as $1 - \sum_{i} p_i^2$, which are minimized in the most homogeneous datasets. The algorithm of DT creation is that at every node, starting with the first one which includes a whole dataset, a split in data is done in order to create two children nodes, each including newly created subsets of the parent node. The best split is decided upon by simply scanning all the features and choosing the best threshold value to maximize the Information Gain ($IG$) defined as \begin{equation}
IG(D_p) = I(D_p) - \frac{N_{left}}{N_p} I(D_{left}) - \frac{N_{right}}{N_p} I(D_{right})\;,
\end{equation}
where $I$ is either entropy or Gini index, $D_p$, $D_{left}$, $D_{right}$ are datasets of a parent, left, and right child node, respectively. A tree can be built as long as there is data belonging to more than one class in leafs which are the last nodes. 

If a single tree was used, such an approach would lead to overfitting. Therefore, trees are usually pruned by setting a maximum level of their depth. Once a tree is created, the inference on a data point is performed by moving along the path of decisions through the tree. Once the leaf corresponding to a given data point is reached, classification probabilities are given by class probabilities calculated during the training procedure in a given leaf. A single DT is a well-working model by itself, however its biggest disadvantage is its limited predictive power due to pruning, which is the only way to regularize the model and prevent overfitting. Many trees are therefore used to create a random forest.

\subsubsection{Bagging}

The method of differentiating DTs in RF is called Bootstrap Aggregation \citep[bagging,][]{Breiman:1996}. Every tree is built using a subsample of data, equally sized as the original dataset, created by uniform sampling with replacement. Features are also sampled, but in classification problems one usually selects $\sim\sqrt{n}$ features, where $n$ is the total number of features. Such an approach maximizes the differences between the trees built in an ensemble. The final decision of an ensemble is made by majority voting, which means that a class that obtained the most votes is chosen as the final answer. In addition, probabilities for the particular classes can be obtained by simply calculating the fraction of the trees which voted for a given class.

Bagging makes RF significantly different from a single DT. Its most important aspect is to introduce a way of regularizing the DTs and preventing overfitting of the final RF output by additional averaging over many simpler classifiers. This means that the DTs no longer have to be pruned, and in fact usually very good results are achieved by fully-built DTs. The last thing to mention is that all the trees are built independently of each other, which prevents overfitting when too many DTs are trained. At some point, all the possible different trees and corresponding decision boundaries have been built and no model improvement is introduced with new trees.

\subsection{Performance evaluation}
\label{subsection_performance_evaluation}

In ML methods it is desirable to separate out validation and test datasets from the training sample, in order to estimate model performance on observations which were not included in model creation. The validation set is used to select the best model and its parameters, while the test set is needed to report final scores and should never be employed to choose algorithm parameters, in order to eliminate the possibility of overfitting to a particular training sample. In practice, if validation and test scores are significantly different, more regularization should be applied to the model. In our application, as the test set we choose a random 20\% subsample of the full training set described in \S\ref{subsection_training_set}. The remaining 80\% of the training data are then used in a five-fold cross-validation procedure, in which they are divided into five separate equally-sized subsets, and four of them are used for the training, while validation scores are calculated on the fifth subsample. The training process is repeated five times, with a different subset used for the validation each time. This gives a total of five values for every metric used, which are then averaged to create the final validation results.
	
As far as the evaluation metrics are concerned, we use several of them in order to  quantify model performance better than would be possible with individual scores. The basic metric used for three-class evaluation is the accuracy, which measures the fraction of correctly classified observations. Additionally, as the main goal of this work is to select quasars, we transform the classification output into a binary one. This is done by simply summing probabilities of stars and galaxies into a new class called \texttt{rest}, and evaluating the performance of the QSO vs. \texttt{rest} problem. To this aim, apart from accuracy applied on the binary problem, we use the purity (precision), completeness (recall), and F1, a harmonic mean of precision and recall. If TP is the number of correctly classified positives, FP the number of incorrectly classified positives, FN the number of incorrectly classified negatives, then the metrics are given by:
\begin{itemize}
	\item[*] $ purity = TP/(TP + FP) $;
	\item[*] $ completeness = TP/(TP + FN) $;
	\item[*] $ F1 = 2 \cdot purity \cdot completeness / (purity + completeness) $.
\end{itemize}
In our case, the positive class consists of quasars, and the negative class of stars plus galaxies. The last binary metric we use is the area under the receiver operating characteristic curve (ROC AUC) based on the output probability for the quasar class only. The ROC curve is created by plotting the true positive rate (TPR) against the false positive rate (FPR) at various probability threshold settings. TPR is the same value as completeness, while FPR is also known as the probability of false alarm and can be calculated as
$ FPR = 1 - specificity = 1 - TN/(TN + FP) $.
The last validation tool used here is the confusion matrix, which shows relations between ground-truth and predicted labels for the multiclass problem. These metrics are used in our machine learning experiments both to select the most appropriate algorithm and set of features.

\subsection{Feature selection} \label{subsection_feature_sel}

Except for deep learning, in which models learn to detect features, in any other machine learning application it is important to properly design the feature set. First, we want to provide as much information to the model as possible, as it will learn which features are really important, and create the discriminative patterns. However, the features must carry useful information to avoid confusing the model and deteriorating its performance. In the worst case scenario, using non-discriminative features can lead to overfitting, which means that the learning patterns work well on training data only, while being inadequate for the general problem. It is thus important to perform a proper experimental analysis and select the features which maximally improve model performance.

In applications with hundreds of features, the process of choosing their best subset can be complex. Here, we use the method of backward elimination \citep{Harrell:2001}. We start with all the available features, even those that, according to our knowledge, may not be very important, and apply a model which calculates feature importance (such as for instance the RF). After initial training, we sort all the features according to their importance and perform iterative removal of the least important ones (in some groups rather than individually). After each removal, we validate the performance of the new model. In this way, within a linear complexity with respect to the number of features, we find the best feature set and optimize model performance.

Feature importance for a single DT is calculated by simply summing all the IGs from a given feature over the whole tree. For the full RF, this value is averaged between all the DTs for each feature. From this, relative importance of features can be calculated as percentages, providing quantitative information on their usefulness in solving a problem.

Machine learning algorithms work more efficiently if they are provided not only with basic features but also their combinations, if those are correlated with the ground truth data like labels in case of a classification. A popular way to extend the feature set is to combine already existing features using simple algebraic operations \citep{Piramuthu:2009}. Colors (differences of magnitudes from various passbands) are one of such ways; another combination popular in ML is feature ratios, and we tested ratios of magnitudes as an extension of the feature space. 

\begin{figure}
	\resizebox{\hsize}{!}{\includegraphics{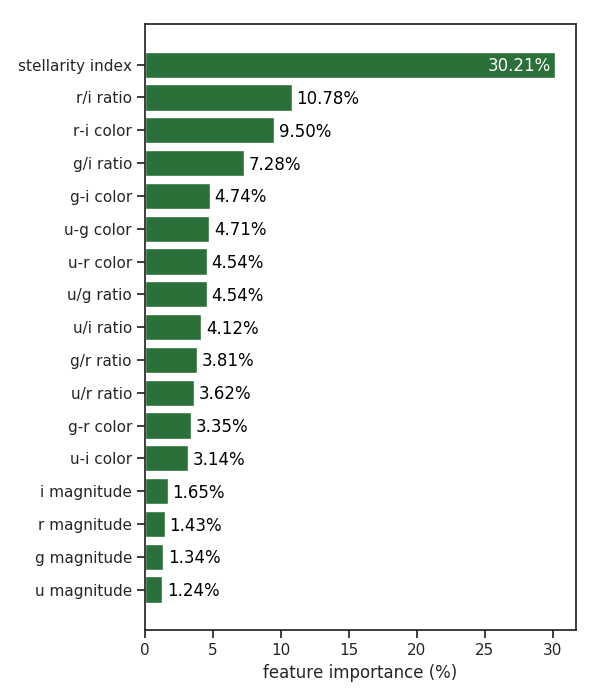}}
	\caption{Features of the final model sorted according to their importance.}
	\label{fig:feature_importance}
\end{figure}

We verified the usefulness of a large number of features from the KiDS DR3 database, such as $ugri$ magnitudes, their differences (colors), their ratios, and also fluxes in all available apertures, observation errors, ellipticity, as well as star/galaxy separators. By applying the above described method of feature importance evaluation, we created the final feature set which provided the best model performance. It consists of 17 features in total: four $ugri$ magnitudes, six resulting colors, six magnitude ratios, and \texttt{CLASS\_STAR}. Figure \ref{fig:feature_importance} quantifies the feature importance in percentages. The stellarity index is a very useful feature, and its role is to provide a nearly perfect separation between galaxies and point-like objects like stars and quasars. Importances of colors and ratios for the same magnitude pairs are similar, which means that they are similarly useful in providing information to the model.

The results illustrated in Fig.~\ref{fig:feature_importance} show that magnitude values are of much less importance for the classification than colors and magnitude ratios. Therefore, in addition to the fiducial approach where all the listed features were used, we have experimented with a classification setup without magnitudes. In such a case, the purity and completeness of the quasar classification measured on the test data were worse by $\sim 1.5\%$, which is mostly due to increased confusion with stars. We also tested the no-magnitude model by generating its predictions on the inference set and comparing the results with those where the whole feature set was used. Only 67\% of quasar candidates identified by the model which includes magnitudes were also classified as QSOs in the \quotes{color-only} case, and almost all of the rest were classified as stars. Based on these findings, we conclude that the model in which magnitudes are not used performs worse, and in particular leads to a higher rate of misclassification with stars. Our default approach is therefore to use all the features shown in Fig.~\ref{fig:feature_importance}.

Figure \ref{fig:feature_importance} indicates also that among the four KiDS DR3 passbands, the $u$ band is of least importance for our classification task. This is also the band which has the largest fraction of missing or excessively noisy observations, removed at the data preparation stage. One could therefore attempt classification based on only $gri$ bands, which would give larger training and inference datasets than in our case (i.e., fewer sources would have been removed in the procedure described in \S\ref{subsub_data_preproc}). In the present application, this would however lead to significant limitation of the feature space, removing in total 7 of the 17 features. We therefore postpone experiments without using the $u$ band to the future KiDS data releases which will incorporate also VIKING NIR photometry and therefore greatly extend the feature space available.

\subsection{Feature space limitation} \label{subsection_feature_space}

In addition to selecting the most relevant features, we need to make sure that the training data cover the feature space sufficiently well for the classification in the inference data to be robust. We already discussed in \S\ref{section_data} that the SDSS training data are much shallower than the full KiDS, therefore in this work the KiDS data is limited to $r<22$ mag to avoid extrapolation. In this subsection we provide details on feature space limitation by analyzing its full multidimensional properties.

One can understand machine learning models as complex decision boundaries in the training feature space. The models are expected to learn true patterns, which should then extend their applicability to new datasets, such as the KiDS inference sample in our case. However, for the points which lie outside of the original region of feature space for which decision boundaries were created, model predictions may implement a classification function extrapolated from the training data, which may then not agree with the patterns outside of the training set. The most straightforward solution is to simply match the inference dataset to the training sample. In our case, the simplest approach is to limit the data to $r<22$ mag. However, one could also work in color space only, without using magnitudes, and perform a cut of $u-g>0$ instead (see right panel of Fig.~\ref{fig:r_feature_space}), which would significantly extend the inference dataset, giving about twice as many entries than the 3.4 million in the fiducial sample limited to $r<22$. Below, in \S\ref{subsection_tsne} we show why a cut in the $r$-band magnitude is more appropriate for our model than a cut in the $u-g$ color.

Cuts performed in single features allow for a better match between the training and inference sets in these particular dimensions. However, as the final classification is performed in a space of significantly larger dimensionality, the usefulness of such an approach is rather limited. A match between individual features does not have to imply proper coverage of the full feature space. A simple counterexample is a 2D square covered by data points drawn from a 2D Gaussian distribution and separated into two subsets by a diagonal. In such case, the histograms of single features show overlap of data in individual dimensions, while in fact there is no data from two subsets overlapping in 2D at all. Therefore, we look in more detail at coverage in the multidimensional feature space of the training and inference data. This is done by projecting the feature space onto two dimensions using the t-SNE method.

\subsubsection{Visualization with t-SNE}
\label{subsection_tsne}

\begin{figure}
	\resizebox{\hsize}{!}{\includegraphics{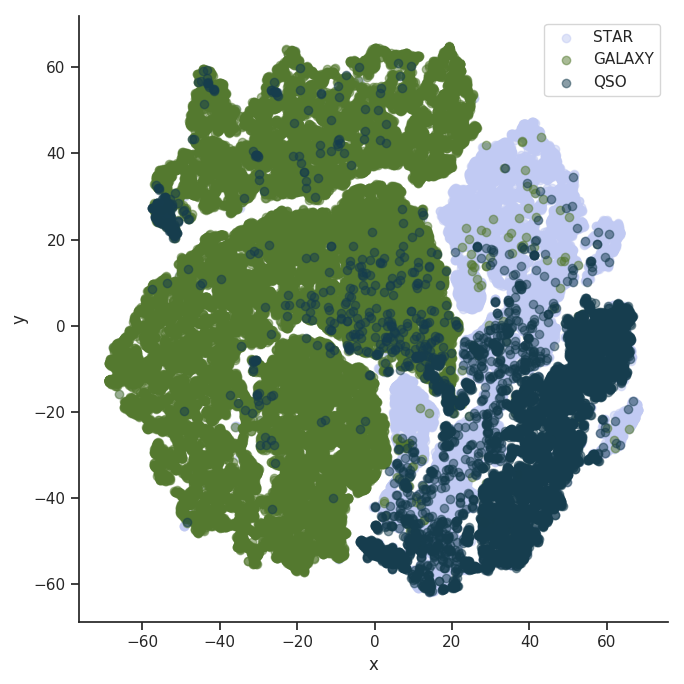}}
	\caption{t-SNE visualization of the training dataset. The plot illustrates a projection of the multidimensional feature space onto a 2D plane, where \textit{x} and \textit{y} are arbitrary dimensions created during the visualization process. Labeled training data are mapped with three different colors as in the legend.}
	\label{tsne_training}
\end{figure}

\begin{figure*}
	\resizebox{\hsize}{!}{\includegraphics{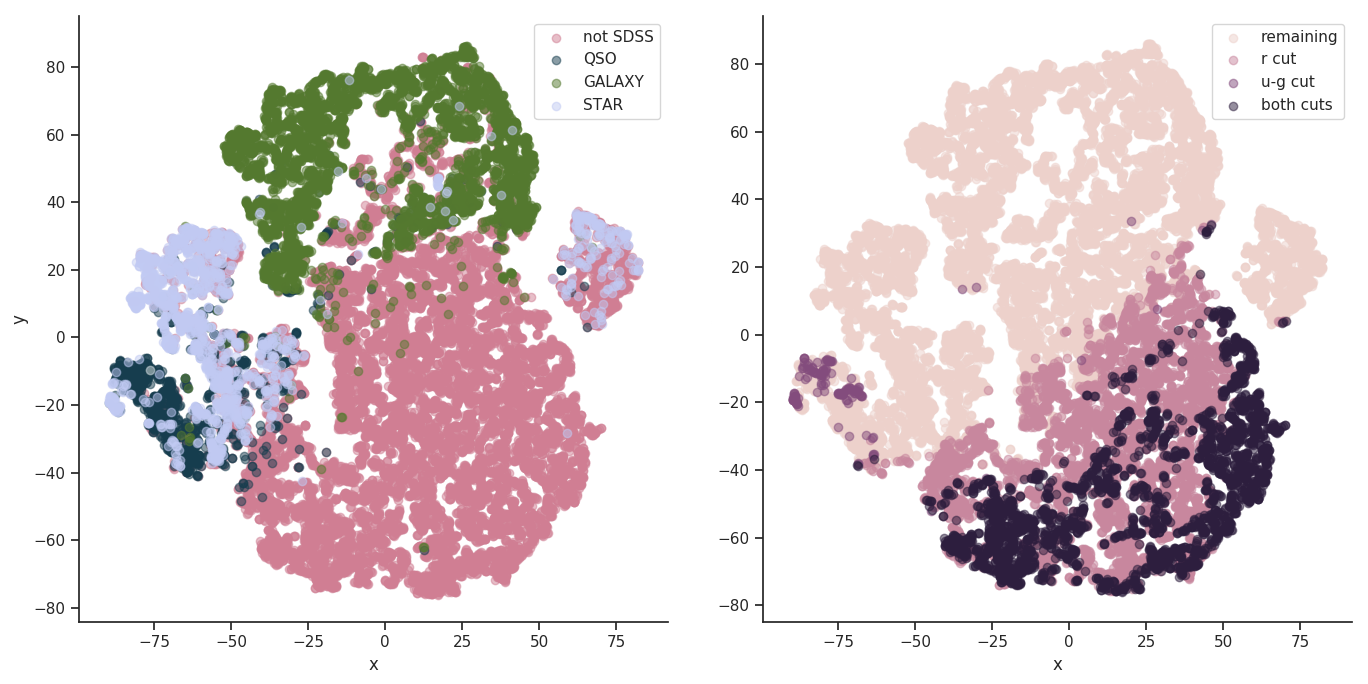}}
	\caption{t-SNE visualization of the catalog feature space. \textit{Left:} datapoints with SDSS labels from the training set (green and light and dark blue) together with those from the inference sample without training coverage (pink). \textit{Right:} results of applying magnitude and color cuts on the inference sample. The darkest color shows objects removed by both $r<22$ and $u-g>0$ criteria simultaneously, colors in between stand for objects removed by only one of those cuts, and the lightest points are left after the cuts.}
	\label{tsne_catalog}
\end{figure*}

There are many ways of mapping $N$-dimensional feature spaces onto 2D projections. A popular one in astronomy is Self Organizing Map \citep[SOM,][]{Kohonen:1997}, and a relevant example of its usage is the mapping of multicolor space to visualize which regions are not covered by spectroscopic redshifts \citep{Masters:2015}. Here, we use an another advanced visualization method, the t-distributed stochastic neighbor embedding \citep[t-SNE,][]{Maaten:2008}, which finds complex nonlinear structures and creates a projection onto an abstract low-dimensional space. Its biggest advantage over other methods is that t-SNE can be used on a feature space of even several thousand dimensions and still create a meaningful 2D embedding. Moreover, unlike in SOM where datapoints are mapped to cells gathering many observations each, in t-SNE every point from the $N$-dimensional feature space is represented as a single point of the low dimensional projection. This makes t-SNE much more precise, allowing it to plot the exact data point values over visualized points as different colors or shapes, making the algorithm output easier to interpret. Some disadvantages of using t-SNE are its relatively long computing time
and its inability to map new sources added to a dataset after the transformation process, without running the algorithm again.

In case of ML methods which use many features at once during the calculations, it is useful to normalize every feature in order to avoid biases related to their very different numerical ranges (such as for magnitudes vs. colors). This does not apply to RF, which uses only one feature in each step of data splitting; however, it does affect the t-SNE algorithm which calculates distances based on all available features. In order not to artificially increase the importance of the features with larger numerical values, we always scale every feature individually to the range $[0, 1]$, as a preprocessing step in the visualization. The transformation of each feature is given by $F_i' = (F_i - F_{min}) / (F_{max} - F_{min})$, where $F$ stands for a given feature, $F_i$ is its value for the $i$-th data point, and $F_{min}$ and $F_{max}$ represent the minimum and maximum values of this feature in a considered dataset.

Our first t-SNE visualization is applied to the training dataset, using the full 17D feature space selected in \S\ref{subsection_feature_sel}, and it gives important information whether the automated quasar detection can be performed at all in the feature space provided by KiDS DR3. As shown in Fig.~\ref{tsne_training}, most of the quasars form their own cluster in the 2D projection, while some do indeed overlap with stars. This does not necessarily mean that those observations are not distinguishable by classification models which work in the original feature space, but it does point at a problem that perhaps additional features should be added, such as magnitudes and colors at other wavelengths than the currently used $ugri$ ones. We will study this issue in the near future with extended KiDS+VIKING data from forthcoming KiDS data releases.

We now turn to a comparison of the training and inference datasets. For that purpose we join the training set with a random subsample of the inference data of similar size as the training (this is to speed up the computation which for the full KiDS DR3 would be very demanding). In Fig.~\ref{tsne_catalog} we show projections of the full 17D feature space (see \S\ref{subsection_feature_sel}) for the dataset constructed this way. Using a new dataset required creating a new visualization, meaning that \textit{x} and \textit{y} axis in this figure are independent of the ones present in the training set visualization (Fig.~\ref{tsne_training}). The left panel includes SDSS labels for the training part of data (green and light and dark blue dots), and \quotes{not SDSS} (pink) standing for inference data which covers feature space outside of the training. Here, the inference data have no magnitude or color cuts applied except for those related to the basic data cleaning (items \#1-\#3 in \S\ref{subsub_data_preproc}). This visualization confirms that a large part of feature space in the inference dataset would not be covered by the training if no additional cuts were applied on the target KiDS sample.

In the right panel of Fig.~\ref{tsne_catalog} we illustrate the effect of magnitude and color cuts on the inference data. The darkest color indicates objects removed by both $r<22$ and $u-g>0$ criteria simultaneously, colors in between show objects clipped by demanding only a single cut, while the lightest points are left after applying any of the cuts. By comparison with the left panel, we clearly see that the $r$-band cut is much more efficient in removing the part of the feature space not covered by training than the cut in $u-g$. It is also worth noting that the color cut would also remove some quasar data points from the feature space covered by training, which is much less the case for the magnitude cut. This is related to the flux-limited character of the SDSS spectroscopic QSOs.

\section{Quasar selection results}
\label{section_evaluation}

In this Section we present and discuss the final quasar selection results in KiDS DR3. By applying the methodology described above, all the sources from our inference dataset of 3.4 million KiDS objects were assigned probabilities of belonging to the three training classes (star, galaxy or quasar). By selecting quasars as those objects which have $p_\mathrm{QSO}>\max(p_\mathrm{star},p_\mathrm{gal})$, we have obtained 192,527 QSO candidates. Here we discuss the properties of this catalog. This is done by first calculating statistical measurements on the test set extracted from the general training sample but not seen by the classification algorithm. The final dataset is also cross-matched with data from the Gaia survey to examine stellar contamination, and with several external quasar catalogs to probe other properties of our sample. Lastly, as a test for completeness, we analyze number counts in the final QSO catalog.

\subsection{Classification results for a test subsample}
\label{subsection_model-testing}

\begin{table}
	\caption{Evaluation metrics for KiDS DR3 calculated from the SDSS-based test set separate from the training and validation data.}
	\centering
	\begin{tabular}{c c c}
		\hline\hline
		classification type & metric & score \\
		\hline
		three-class & accuracy & 96.6\% \\
		\hline
		\multirow{5}{*}{QSO vs. rest} & accuracy & 97.0\% \\
		& ROC AUC & 98.5\% \\
		& purity & 90.8\% \\
		& completeness & 86.6\% \\
		& F1 & 88.7\% \\
		\hline
	\end{tabular}
	\label{table:test}
\end{table}
	
\begin{figure}
	\resizebox{\hsize}{!}{\includegraphics{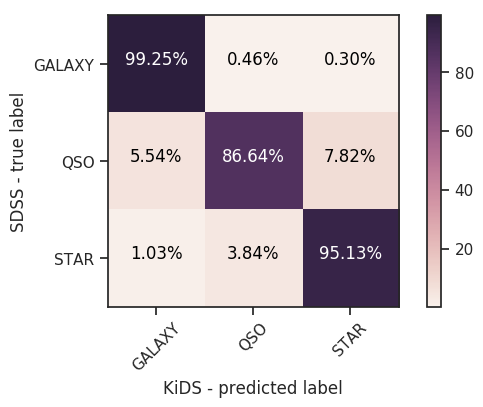}}
	\caption{Normalized confusion matrix of the KiDS DR3 classification calculated for the SDSS test sample.}
	\label{confusion_matrix}
\end{figure}
	
\begin{figure}
	\resizebox{\hsize}{!}{\includegraphics{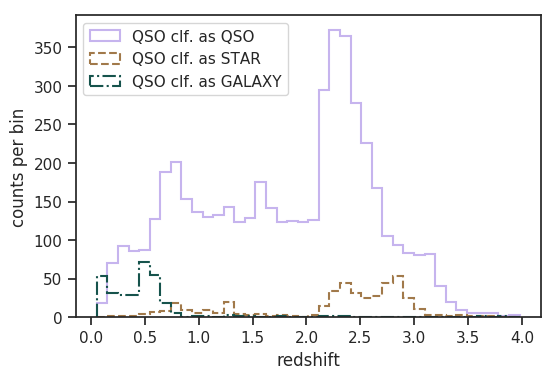}}
	\caption{Redshift distribution of SDSS quasars present also in our KiDS inference sample, separated into the classes predicted by our model. The solid purple line shows correctly classified QSOs, while the orange dashed and green dot-dashed are for true quasars misclassified as stars and galaxies respectively.}
	\label{qso_z}
\end{figure}

The results of our classification were tested as described in \S\ref{subsection_performance_evaluation}. As far as the evaluation metrics are concerned, we measure the accuracy for the three- and two-class (QSO vs. rest) problems, while the ROC AUC, precision, completeness and F1 are provided only for the binary case. All the scores are listed in Table \ref{table:test}. Accuracy of the algorithm is very similar in both two- and three-class cases and amounts to almost 97\%. The ROC AUC provides a very high value of $\sim99\%$. Purity of the final catalog is estimated to be $\sim91\%$, as the quasar test sample is contaminated with ~$\sim7\%$ stars and $\sim2\%$ galaxies. The completeness is a little bit lower, $\sim87\%$. The $F1$ measure gives then $\sim89\%$.

Figure \ref{confusion_matrix} visualizes the classification results in the form of a normalized confusion matrix (CM), from which more information can be extracted. In the case of a normalized CM, cell values are given as percentages which sum up to 100\% in each row. This gives completeness values on the diagonal for each of the classes. From the top row, it is clear that almost all the galaxies are classified correctly, and in the case of stars, a small fraction is misclassified as quasars, which reduces the purity of the QSO catalog. Quasars themselves are the most prone to misclassification, as about 13\% of them are assigned either star or galaxy labels, which translates to the incompleteness of final QSO sample.

In order to better understand the reasons for misclassification, we examine the redshift ranges at which the quasars are assigned incorrect classes. Figure \ref{qso_z} compares redshift distributions of the true SDSS quasars which were classified as QSOs (solid purple), or misclassified as stars (dashed orange) or as galaxies (dash-dotted green). As expected, the QSO-galaxy mismatch happens predominantly at very low redshifts, where the QSO host galaxies can have large apparent size and flux. The flux-limited nature of the spectroscopic training quasars means that at low redshifts, intrinsically less luminous QSOs are included, and the flux of the host galaxy can be comparable or even dominant over that from the AGN. An additional possible explanation is that although we do not explicitly use redshifts in the classification, fluxes and colors of galaxies and quasars are correlated with redshift, so the classification model indirectly learns this relation. Therefore, as the training data are dominated at low redshift by galaxies, this can be problematic for the model.

For better insight into the galaxy/QSO mismatch, we have inspected spectra of 100 randomly chosen sources which were labeled as QSOs by SDSS but classified as galaxies by our algorithm. These objects show signs of AGN emission in the spectrum (broadened lines and prominent high ionization lines), however also the D4000 break and the calcium doublet are visible, characteristic of older stellar populations in the host galaxy. As our classification scheme does not use spectra, the shape of the continuum plays a crucial role in the performance. For that reason, when the emission of the host galaxy is detectable, the SDSS AGNs are often treated by the algorithm as galaxies, regardless of clear presence of AGN signatures in the spectrum.
 
Regarding QSOs incorrectly assigned with a star label, this happens at specific redshift ranges, such as $2.2<z<3.0$, where it is difficult to distinguish quasars from stars with spectral types spanning from late $A$ to early $F$ using broad-band optical filters \citep{Richards:2002,Richards:2009a}. This is exactly the redshift range where we observe the most of the star-QSO misclassification.

The above analysis concerned  the completeness of final QSO sample as a function of redshift. At present we cannot examine a relation between purity and redshift, as we would have to know the redshifts assigned to quasar candidates, including those which in fact are stars or galaxies. As already mentioned, presently in KiDS DR3 the quasars do not have robust photo-$z$s. We plan to address this problem in future studies where both QSO detection and redshift estimation will be performed (see e.g., \citealt{Fotopoulou:2018}). However, for the redshift estimation to be robust, additional near-IR data will be needed. This will be available for KiDS sources starting from DR4 as a result of in-house processing of overlapping VIKING data.

\subsection{Catalog validation with Gaia parallaxes and proper motions} \label{subsection_gaia_motion}

We validate the purity of our KiDS QSO catalog by analyzing parallaxes and proper motions of the contained sources. For this we use the second data release of the Gaia survey \citep{Gaia:2018}, which is currently mapping the entire sky, focusing on stars in the Milky Way, but detecting also extragalactic objects like quasars \citep{Gaia:2016}. At present, over 1.3 billion Gaia-detected sources in the magnitude range $3<G<21$ have measurements of parallaxes and proper motions, therefore, a cross-match with that dataset can be used to test statistically if our QSO candidates are indeed extragalactic. In particular, the QSO candidates are expected to have negligible parallaxes and proper motions in the absence of systematics.

As Gaia is significantly shallower than KiDS ($G<21$ corresponds to $r\lesssim 20$), practically all of the sources from Gaia over the common sky area have a counterpart in KiDS. The reverse of course does not hold, especially since Gaia does not store measurements of extended sources, and in particular only 32\% of our inference sample is also matched to Gaia within $1"$ radius.
In addition, due to considerable measurement errors in source motions at the faint end of Gaia, the test presented here cannot provide an unambiguous star/quasar division for our full inference sample. Moreover, as discussed in detail by \cite{Lindegren:2018}, the measurements of motions in Gaia DR2 have some non-negligible systematics. In particular, even stationary quasars have appreciable scatter in their measured parallaxes and proper motions. A special procedure is therefore needed to validate the contents of our quasar catalog using Gaia, as described below.

In order to analyze the systematics in Gaia DR2 parallaxes and proper motions, \cite{Lindegren:2018} used a sample of quasars, which define a celestial reference frame, known as Gaia-CRF2 \citep{Gaia:2018b}, nominally aligned with the extragalactic International Celestial Reference System and non-rotating with respect to a distant universe. This allowed them to design a set of criteria applied to Gaia measurements to make sure that the selected sources are indeed stationary. As a result, \cite{Lindegren:2018} determined a global mean offset in Gaia parallaxes of $-0.029$ mas. Detecting appreciably higher offsets in the parallax distribution for sources assumed to be quasars would then point to stellar contamination.

\begin{table}
	\caption{Mean values of parallax ($\varpi$), right ascension and declination proper motions (${\mu}_{\alpha*}$ and ${\mu}_{\delta}$), all in milli-arcsecond units, as derived from the Gaia high precision sample (see text for details). First three sets of rows show results for the ground-truth SDSS and KiDS$\times$SDSS training objects. Next, acceptable quasar offsets based on model testing results are presented, while the last two rows show values for the KiDS quasar catalog and its probability-limited subset.}
	\centering
	\begin{tabular}{c c c c c c}
		\hline\hline
		&  & size & $\varpi$ & ${\mu}_{\alpha*}$ & ${\mu}_{\delta}$ \\
		\hline
		\multirow{2}{*}{QSO} & SDSS & 138k & -0.02 & -0.02 & -0.03 \\
		& train & 2.1k & -0.01 & -0.02 & -0.01 \\
		\hline
		\multirow{2}{*}{star} & SDSS & 560k & 0.71 & -1.81 & -6.37 \\
		& train & 7.3k & 0.57 & -6.12 & -6.01 \\
		\hline
		\multirow{2}{*}{galaxy} & SDSS & 3.8k & 0.16 & -0.70 & -2.50 \\
		& train & 78 & 0.29 & -3.60 & -2.93 \\
		\hline
		\multirow{2}{*}{acceptable} & SDSS & - & 0.04 & -0.14 & -0.50 \\
		& train & - & 0.05 & -0.50 & -0.48 \\
		\hline
		\multirow{2}{*}{QSO} & KiDS & 7.1k & 0.21 & -0.27 & -1.08 \\
		& p > 0.8 & 5.8k & 0.09 & 0.14 & -0.42 \\
		\hline
	\end{tabular}
	\label{table:gaia_mean}
\end{table}

\begin{figure*}
	\resizebox{\hsize}{!}{\includegraphics{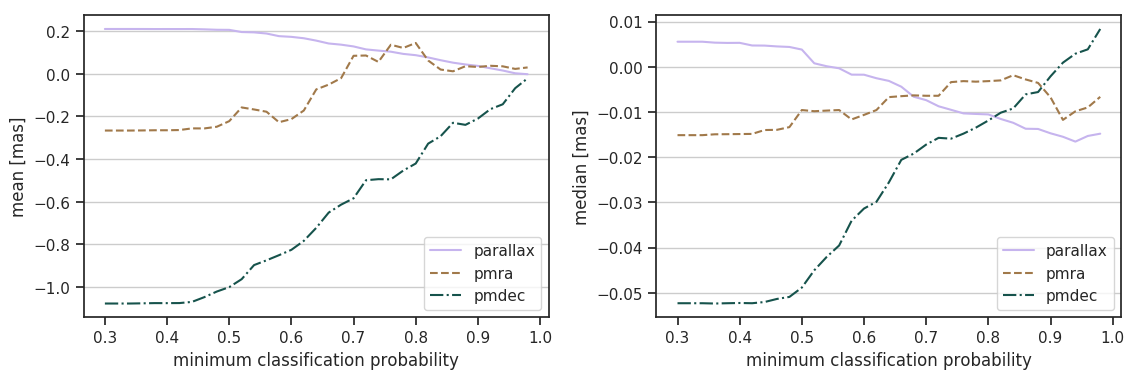}}
	\caption{Mean (left panel) and median (right panel) of parallax (solid purple), right ascension (dashed orange) and declination (dash-dotted green) proper motions, derived from the Gaia high precision sample for KiDS quasar candidates, as a function of minimum quasar probability limit.}
	\label{fig:gaia_mean_sigma_proba}
\end{figure*}

A cross-match of our inference catalog with Gaia DR2 gave almost 1.1 million common objects. Among these, 1 million were classified by the model as stars, 40k as galaxies, and 38k as quasars. As our goal is to evaluate the stellar contamination of the quasar catalog, we cannot directly apply the criteria of Gaia data cleaning from \cite{Lindegren:2018}, as the aim there was to reduce this contamination in a QSO sample. Instead, we define a Gaia high precision sample, by taking sources with parallax and proper motion errors smaller than 1 mas, and compare the measurements for the KiDS sources to those obtained in the same way from the SDSS and KiDS$\times$SDSS training sample (where for the SDSS case we used the full DR14 spectroscopic dataset cross-matched with Gaia DR2). This reduces the number of quasars found in both KiDS and Gaia from  38k to 7.1k.

In order to properly measure the purity of quasar candidates, we have to take into consideration the test results (\S\ref{subsection_model-testing}), according to which our QSO candidate catalog consists of 91\% quasars, 7\% stars and 2\% galaxies. Therefore, we calculate \quotes{acceptable} parallax and proper motion offsets by taking a weighted mean of the respective astrometric quantities with weights of 0.91, 0.07 and 0.02 for ground truth quasars, stars and galaxies. Table \ref{table:gaia_mean} shows parallax and proper motion mean values for ground truth SDSS and training objects, together with the acceptable offsets and values derived for the KiDS quasar candidates. The acceptable offset for parallax ($\varpi$) is about 0.05 mas for both the full SDSS and training QSO samples, and $\sim -0.50$~mas for the proper motion in declination (${\mu}_{\delta}$). The right ascension proper motion (${\mu}_{\alpha*}$) shows inconsistent results between the full SDSS QSO and training datasets. In addition, it varies much more than $\varpi$ and ${\mu}_{\delta}$, and its mean even changes sign depending on the QSO threshold probability, as shown below in Fig.~\ref{fig:gaia_mean_sigma_proba}.

The full catalog of KiDS quasar candidates matched with Gaia shows mean offsets significantly higher than the acceptable levels, which must be an imprint of residual stellar and galaxy contamination. We note however that as we use unclipped means, a fraction of significant outliers can highly influence the means.
Still, those measurements can be used to purify the catalog by limiting the quasars to higher probability values according to our classification model. This makes sense from an ML point of view, as our model was optimized for the training dataset whose properties may differ from the final inference sample. Moreover, we know that some quasars are not easily distinguishable from stars in the optical bands used here, and the two-classes may occasionally overlap in terms of their positions in the feature space (Fig.~\ref{tsne_catalog}). Such objects may have lower classification probability as they are surrounded by sources from an opposite class. This fact can be used to reduce the problem of stellar contamination by simply applying a limit on quasar probability. As shown in Table \ref{table:gaia_mean}, at $p_\mathrm{QSO} > 0.8$ we obtain an acceptable offset for the mean value of ${\mu}_{\delta}$, and close to acceptable for $\varpi$. The absolute value of ${\mu}_{\alpha*}$ is also at an acceptable level for this probability limit, and in fact its mean oscillates around 0 for $p_\mathrm{QSO} \gtrsim 0.7$ (Fig.~\ref{fig:gaia_mean_sigma_proba}).

Figure \ref{fig:gaia_mean_sigma_proba} shows how the mean and median values of parallax and proper motions change as we increase the QSO probability limit. Mean values converge to 0 mas, while median values at this level of precision are required to stay within the QSO mean offsets shown in the first row of Table \ref{table:gaia_mean}. Both mean and median values of the astrometric measurements decrease (in terms of their absolute values) for $p_\mathrm{QSO}>0.5$. For higher QSO probability levels of 0.7 -- 0.8 they are sufficiently close to the acceptable offsets for mean measurements, or 0 mas in case of median values, that at these $p_\mathrm{QSO}$ the quasar candidates can be considered reliable. An exception is the parallax, whose median  changes sign at $p_\mathrm{QSO}\sim0.5$ and continues decreasing to almost $-0.02$~mas at $p_\mathrm{QSO}\sim1$. This is however expected from the offset calculated by \cite{Lindegren:2018} which equals $-0.029$ mas for parallax measurements. 

We consider these results a strong success of our model, especially since this analysis of KiDS objects which are also present in Gaia focuses on the star/quasar separation, which is the most difficult task to solve. Moreover, Gaia measurements may be strongly contaminated with large positive or negative values, resulting from an inconsistent matching of the observations to different physical sources. This may especially affect quasar measurements which require higher resolution than other objects, and therefore can show larger offsets in our catalog than in the training data. Based on this analysis, we suggest to limit the catalog to a minimum classification probability of $p > 0.8$ which favors purity and gives a sample of $\sim$75k quasars, or use a cut of $p > 0.7$ which gives better completeness for a sample of $\sim$100k quasars.

\subsection{Comparison with external quasar catalogs}

{Another method of examining the properties of our quasar catalog is by matching it with other QSO datasets overlapping with KiDS DR3. We use four external samples for this purpose: the spectroscopic 2QZ and 6QZ \citep[][hereafter 2QZ]{Croom:2004}, and three photometric samples \citep[][hereafter R09, R15 and DP15 respectively]{Richards:2009a,Richards:2015,DiPompeo:2015}. 2QZ includes confirmed quasars, stars and galaxies, while the photometric catalogs are probabilistic, based on selection from SDSS (R09) and SDSS+WISE (R15 \& DP15). In addition, only 2QZ significantly overlaps with the KiDS footprint, while the others (R09, R15 and DP15) cover the SDSS area\footnote{Another QSO sample that could be used is 2SLAQ \citep{Croom:2009} but it has much less overlap with KiDS DR3 than those considered here.}. They also have different depths, as illustrated in Fig.~\ref{fig:ext_qso_hist} which shows $r$-band distributions of cross-matches between the full KiDS DR3 and the four discussed catalogs. Among these, 2QZ is considerably shallower ($r<21$) than our inference dataset, which is the main reason why we have not included it in our training set. As a result, the number of cross-matches between our inference catalog and the external datasets is not expected to be very large. Indeed, taking from the comparison catalogs sources which are labeled as quasars, we find respectively 5.4k objects of our inference sample in 2QZ, 17k in R09, 18k in R15 and 43k in DP15. Of these, respectively 5.2k (97\%), 14.6k (86\%), 16.4k (91\%) and 31.8k (74\%) have quasar labels in our catalog (see Table \ref{table:external_qso}). A relatively lower consistency between our QSOs and DP15 might be related to the fact that in the DP15 some quasar candidates have probabilities as low as $p_\mathrm{QSO}>0.2$. It should be stressed that the probabilistic character of the R09, R15 and DP15 datasets means that they can be only used for qualitative rather than quantitative comparisons. Unlike the spectroscopic 2QZ, these 3 photometric QSO datasets cannot be treated as ground truth and we will use them mainly to test the consistency between our model and the external approaches, and to further validate the minimum QSO probability at which our quasar catalog is robust.

\begin{figure}
	\resizebox{\hsize}{!}{\includegraphics{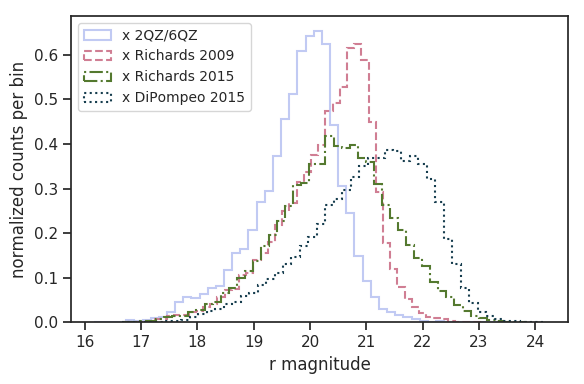}}
	\caption{Normalized distributions of the $r$-band magnitude for cross-matches between the KiDS DR3 and four overlapping quasar catalogs, as indicated in the legend.}
	\label{fig:ext_qso_hist}
\end{figure}

\begin{table}
	\caption{Contributions of the classes predicted by our model, starting with the whole inference dataset and then moving to its cross-matches with external quasar catalogs: one spectroscopic (2QZ), providing ground truth, and 3 photometric, which are probabilistic. In those cross-matches, the highest quasar contribution is expected.}
	\centering
	\begin{tabular}{l c c c c}
		\hline\hline
		& size & star & quasar & galaxy \\
		\hline
		KiDS DR3 inference dataset & 3.4M & 35\% & 6\% & 59\% \\
		\hline
		$\times$ 2QZ and 6QZ  & 5.4k & 2\% & 97\% & 1\% \\
		$\times$ \cite{Richards:2009a} & 17k & 9\% & 86\% & 5\% \\
		$\times$ \cite{Richards:2015} & 18k & 6\% & 91\% & 3\% \\
		$\times$ \cite{DiPompeo:2015} & 43k & 15\% & 74\% & 11\% \\
		\hline
	\end{tabular}
	\label{table:external_qso}
\end{table}

The availability of three-class spectroscopic labels in 2QZ allows us to calculate the same metrics as for the SDSS test sample discussed in \S\ref{subsection_model-testing}. The cross-match with KiDS reduces 2QZ to 7.8k objects which, in terms of spectroscopic 2QZ labels, consists of 5.4k QSOs, 2.4k stars and only 15 galaxies. We obtain high metric values in this case: three-class accuracy of 95\%, QSO purity of 95\% and completeness of 97\%. This is summarized in Table \ref{table:2qz}, and Fig.~\ref{conf_mx_2qz} which shows the relevant confusion matrix. These results give an independent confirmation of the very good performance of our QSO classification also at the bright end, here evaluated on a truly \quotes{blind} test set which was not part of the general training data. In particular, that comparison sample had been preselected from different input imaging and according to different criteria than SDSS, although we note that some 2QZ quasars are now included in the SDSS database.

\begin{table}
	\caption{Evaluation metrics for KiDS DR3 quasar classification, calculated from the 2QZ test set.}
	\centering
	\begin{tabular}{c c c}
		\hline\hline
		classification type & metric & score \\
		\hline
		three-class & accuracy & 94.5\% \\
		\hline
		\multirow{5}{*}{QSO vs. rest} & accuracy & 94.9\% \\
		& ROC AUC & 96.4\% \\
		& purity & 95.3\% \\
		& completeness & 97.4\% \\
		& F1 & 96.3\% \\
		\hline
	\end{tabular}
	\label{table:2qz}
\end{table}

\begin{figure}
	\resizebox{\hsize}{!}{\includegraphics{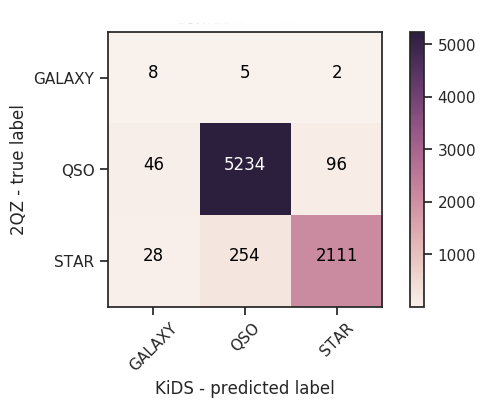}}
	\caption{Confusion matrix of the KiDS DR3 classification calculated for the overlapping 2QZ sources.}
	\label{conf_mx_2qz}
\end{figure}

\begin{figure}
	\resizebox{\hsize}{!}{\includegraphics{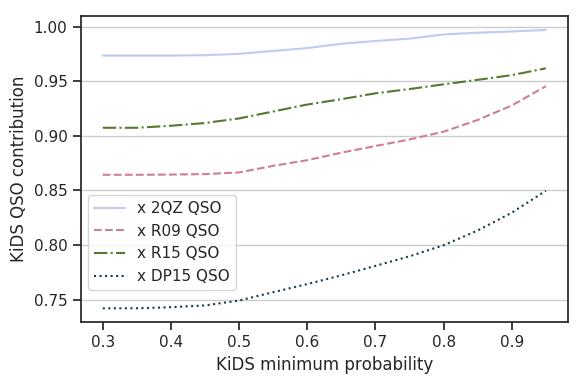}}
	\caption{Proportion of KiDS QSOs in cross-matches with external quasar samples as a function of KiDS minimal classification probability. See text for details of the datasets.}
	\label{ext_qso_proba}
\end{figure}

As already mentioned, the remaining QSO catalogs used here for validation are probabilistic, therefore matching with them leads to more qualitative than quantitative evaluation. Still, we observe good consistency between our detections and those external models, especially for the R09 and R15 cases.
Together with 2QZ, we use the overlapping QSO samples to further improve the purity of our quasar catalog. For this, we employ the probabilities delivered by the RF model, as was already discussed in \S\ref{subsection_gaia_motion}. Except for the 2QZ case, this serves mostly to improve the consistency between our model and those external methods. Figure \ref{ext_qso_proba} shows how the consistency between the models rises as we increase the minimum probability at which we accept the KiDS QSO classification. For 2QZ, we see  excellent consistency for all the probability values and almost perfect (i.e., KiDS QSO contribution $\sim1$) for $p_\mathrm{QSO} > 0.8$. For the probabilistic catalogs, we observe that many external quasars are classified by our method with $p_\mathrm{QSO} > 0.8$, which we deduce from the increase in consistency above this value. Those results fully agree with the conclusion from \S\ref{subsection_gaia_motion} which states that it is a good option to limit our identifications to $p_\mathrm{QSO} > 0.8$ when optimizing the purity of the quasar catalog.

\subsection{Validation using WISE photometric data}

We also validate our KiDS QSO catalog using mid-IR data from the full-sky Wide-field Infrared Survey Explorer \citep[WISE,][]{Wright:2010}. Despite being relatively shallow ($\sim17$ (Vega) in the 3.4 $\mu$m channel), WISE is very efficient in detecting quasars at various redshifts. In particular, QSOs in WISE stand out having very \quotes{red} mid-IR $W1-W2$ ($[3.4 \mu\mathrm{m}] - [4.6 \mu \mathrm{m}]$) color \citep[e.g.,][]{Wright:2010,Jarrett:2011,Jarrett:2017}. The general rule-of-thumb for QSO selection in WISE is $W1-W2>0.8$ \citep{Stern:2012}, but more refined criteria are needed to obtain pure and complete quasar samples from WISE \citep[e.g.,][]{Assef:2013,Assef:2018}. In particular, a non-negligible number of optically selected QSOs have $W1-W2$ significantly lower than the 0.8 limit \citep[e.g.,][]{Kurcz:2016}. That being said, QSOs are generally well separated from galaxies and stars in the $W1-W2$ color with some minimal overlap for $W1-W2<0.5$.

Although there exist QSO or AGN catalogs selected from WISE only \citep{Secrest:2015,Assef:2018}, here we use the entire AllWISE data release \citep{Cutri:2013} for the cross-match, as our goal is to derive the mid-IR $W1-W2$ color of all the KiDS quasar candidates. We have cross-matched both our training set and the output catalog with AllWISE using a $2"$ matching radius (a compromise between KiDS sub-arcsecond resolution and the $\sim6"$ PSF of WISE). We first note that $\sim81\%$ of our training set have counterparts in AllWISE, while for the inference sample this percentage is lower, $\sim45\%$, mostly due to WISE being considerably shallower than KiDS in general. We also confirm the observation from \cite{Kurcz:2016} that a large fraction of SDSS-selected quasars have $W1-W2<0.8$ ($\sim22\%$ in the cross-match of our training set quasars with WISE detections).

For the output catalog, the distribution of the $W1-W2$ color for QSO candidates in the matched sample is in good agreement with that of the training set, with a slight preference to \quotes{bluer} colors which might actually reflect true properties of these optically-selected quasars rather than problems with our algorithm. Interestingly, this distribution shifts towards redder values of $W1-W2$ when cuts on higher pQSO are applied. This is illustrated in Fig.~\ref{fig:w1-w2}, which shows that for $p_\mathrm{QSO}>0.9$, the distribution of $W1-W2$ for the KiDS QSO candidates is very similar to that of the SDSS spectroscopic quasars matched to WISE. This is remarkable given that nowhere in our classification procedure any mid-IR information was used, which additionally confirms the purity of KiDS quasar catalog.

\begin{figure}
	\resizebox{\hsize}{!}{\includegraphics{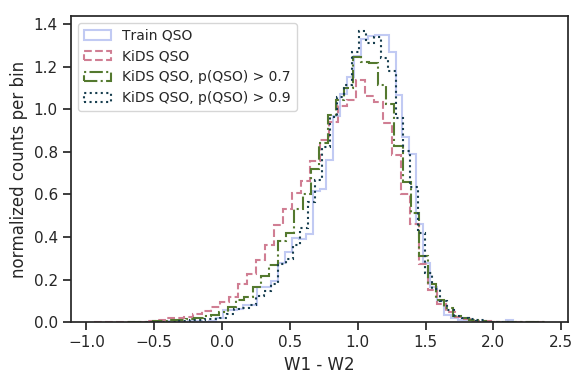}}
	\caption{Distribution of the mid-infrared $W1-W2$ color (3.4 $\mu$m - 4.6 $\mu$m, Vega) of quasar candidates in our KiDS sample, derived from a cross-match with all-sky WISE data. We show histograms for all KiDS QSOs matched with WISE, as well as for two examples of the probability cut: $p_\mathrm{QSO} > 0.7$, which is recommended to increase the purity of the sample, and $p_\mathrm{QSO} > 0.9$ to illustrate how the resulting $W1-W2$ changes when the minimum probability considerably increases.}
	\label{fig:w1-w2}
\end{figure}

\subsection{Number count analysis}

\begin{figure}
	\resizebox{\hsize}{!}{\includegraphics{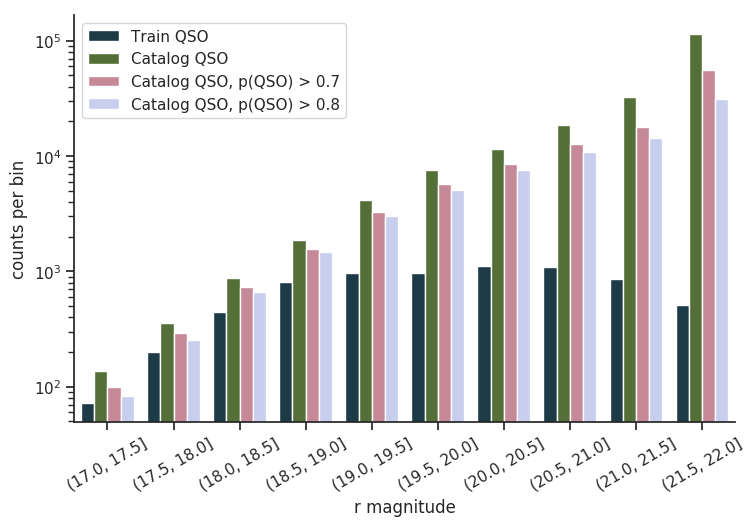}}
	\caption{Number counts of SDSS quasars and KiDS quasars classified in this paper. 
Together with the full QSO candidate sample, we also show samples limited to quasar probabilities above 0.7 and 0.8, which are the cuts we suggest applying to improve the purity of the sample.}
	\label{number_count}
\end{figure}

As another test of completeness, we now compare the number counts of the quasars used for training and those in our final sample. This is done for the $r$ band on which KiDS detections in the multiband catalog are based. The SDSS DR14 quasars used here as the training sample were preselected based on several color cuts and other criteria \citep[e.g.,][]{Blanton:2017}, which results in various levels of incompleteness as a function of redshift and magnitude. On the other hand, a QSO sample selected from imaging data as the one resulting from our work is expected to provide a much more complete sample, ideally volume-limited. In Fig.~\ref{number_count} we compare $N(r)$ for the input SDSS spectroscopic quasars with results from our classification. In the latter case, we show number counts for the general QSO candidate sample and for the cut at $p_\mathrm{QSO} > 0.7$ and $p_\mathrm{QSO} > 0.8$, determined above as optimal for improving the purity of this dataset.

The counts shown give the total number of sources per bin for the full respective samples, that is not normalized by area. Therefore, the comparison has mostly qualitative character. It serves as a verification that the number counts in our photometrically-selected quasar sample steadily rise up to the limiting magnitude of the sample, also for the cases of minimum probability thresholds applied. In other words, the incompleteness visible for the SDSS training sample at the faint end is not propagated to the final selection of KiDS QSOs. 

\section{Conclusions and future prospects}
\label{section_conclusions}

We have presented a photometry-based machine learning selection of quasars from optical KiDS DR3 $ugri$ data. This is the first such comprehensive study using this dataset, serving as a step towards future scientific analyses employing these objects. For our selection we employed the random forest supervised learning algorithm, using the spectroscopic SDSS DR14 data overlapping with KiDS as the training set. We put particular emphasis on choosing the parameter space used for the classification, by examining the importance of various photometric features available from the KiDS database. We eventually decided on a 17 dimensional feature space, including magnitudes, their ratios, colors, and the stellarity index. We also verified that a classifier which does not use magnitudes directly performs worse, showing increased confusion between quasars and stars.

Applying t-SNE, an advanced tool to project multidimensional data onto 2D planes, we examined how the feature space is covered by the training data. This was necessary to appropriately limit the target (inference) sample in order to avoid extrapolation, the reliability of which is difficult to establish in case of supervised learning approaches. In particular, the currently available training data from SDSS did not allow us to probe beyond the limit of $r=22$ mag, which is 3 mag shallower than KiDS $5\sigma$ depth. For this limitation to be overcome, future and deeper QSO training data will be needed, and indeed these should be available from such spectroscopic campaigns overlapping with the KiDS footprint as SDSS eBOSS, DESI, or 4MOST.

The random forest classifier identified about 190,000 quasar candidates among the 3.4 million objects in our KiDS DR3 inference sample. We validated the purity and completeness of this catalog both on a test sample extracted from the SDSS spectroscopic data, as well as on external datasets, such as Gaia, 2QZ, WISE, and photometric QSO catalogs derived from SDSS and WISE. All these tests, together with a number count analysis of the output QSO catalog, indicated high levels of purity and completeness (respectively 91\% and 87\% for the test sample). The main contaminants are stars, while incompleteness seems localized to specific redshifts at which stars and quasars overlap in the $ugri$ color space, or the emission of the AGN host galaxy misleads the classifier which is not provided with all the spectral features. According to the analysis made with Gaia, for scientific usefulness the quasar sample should be limited to a minimum probability of $p_\mathrm{QSO} > [0.7, 0.8]$, where the lower end favors completeness and the higher end improves purity. Those samples include 100k and 75k quasar candidates, respectively.

The catalog is released publicly at \url{http://kids.strw.leidenuniv.nl/DR3/quasarcatalog.php}, and it consists of KiDS objects which were subject to our model classification (the inference set). Apart from basic columns present in KiDS DR3, we add the resulting probabilities for each class in \texttt{QSO}, \texttt{STAR} and \texttt{GALAXY} columns. The final classification, given by the class with the highest probability, is given in the \texttt{CLASS} column. To obtain the full catalog of all the 190k quasars candidates, one has to query $\mathtt{CLASS} == "QSO"$, while the high precision catalog is accessible by taking $\mathtt{QSO} > 0.8$. The code, written in Python and Jupyter Notebook, is shared publicly at \url{https://github.com/snakoneczny/kids-quasars}. The code is meant to be self explanatory, but in case of any questions please do not hesitate to contact the main author.

Starting from DR4, future KiDS data releases will incorporate not only the optical but also near-IR photometry from the VIKING survey. This will provide nine band magnitude space spanning from the $u$ band up to $K_s$ of 2.2 $\mu$m. Availability of these longer-wavelength data is expected to improve quasar detection \citep[e.g.,][]{Peth:2011,Maddox:2012}, and in a forthcoming study we will present supervised QSO classification applied to KiDS DR4 + VIKING data. A much larger feature space will also allow us to test classification on objects removed from the present catalog due to missing features, which should increase the number of classified objects and the completeness of the resulting quasar catalog. Another important aspect, essential for the full scientific usefulness of photometrically identified quasars, is to estimate their redshifts. We plan to work on this in the near future, using the nine band data and potentially adding also information from the WISE all-sky survey.

\begin{acknowledgements}

Based on data products from observations made with ESO Telescopes at the La Silla Paranal Observatory under programme IDs 177.A-3016, 177.A-3017 and 177.A-3018, and on data products produced by Target/OmegaCEN, INAF-OACN, INAF-OAPD and the KiDS production team, on behalf of the KiDS consortium. OmegaCEN and the KiDS production team acknowledge support by NOVA and NWO-M grants. Members of INAF-OAPD and INAF-OACN also acknowledge the support from the Department of Physics \& Astronomy of the University of Padova, and of the Department of Physics of Univ. Federico II (Naples).\\

Funding for SDSS-III was provided by the Alfred P. Sloan Foundation, the Participating Institutions, the National Science Foundation, and the U.S. Department of Energy Office of Science. The SDSS-III website is \url{http://www.sdss3.org/}. SDSS-III is managed by the Astrophysical Research Consortium for the Participating Institutions of the SDSS-III Collaboration including the University of Arizona, the Brazilian Participation Group, Brookhaven National Laboratory, Carnegie Mellon University, University of Florida, the French Participation Group, the German Participation Group, Harvard University, the Instituto de Astrofisica de Canarias, the Michigan State/Notre Dame/JINA Participation Group, Johns Hopkins University, Lawrence Berkeley National Laboratory, Max Planck Institute for Astrophysics, Max Planck Institute for Extraterrestrial Physics, New Mexico State University, New York University, Ohio State University, Pennsylvania State University, University of Portsmouth, Princeton University, the Spanish Participation Group, University of Tokyo, University of Utah, Vanderbilt University, University of Virginia, University of Washington, and Yale University.\\

SN, MBi, AS \& AP are grateful for support from Polish Ministry of Science and Higher Education, MNiSW, through a grant DIR/WK/2018/12.
SN was supported by the MNiSW through grant number 212727/E-78/M/2018.
MBi was supported by the Netherlands Organization for Scientific Research, NWO, through grant number 614.001.451. 
AP was supported by National Science Centre, Poland, through grant No. 2017/26/A/ST9/00756.
CS has received funding from the European Union’s Horizon 2020 research and innovation programme under the Marie Sklodowska-Curie actions grant agreement No 664931.
MBr acknowledges the INAF PRIN-SKA 2017 program 1.05.01.88.04 and the funding from MIUR Premiale 2016: MITIC.
NRN acknowledges support from the European Union’s Horizon 2020 Sundial Innovative Training Network, grant n.721463, and from the 100 Top Talent Program of the Sun Yat-sen University, Guandong Province.
\\

The following software was used: TOPCAT \citep{Topcat:2005}, Python programming language (Python Software Foundation, \url{https://www.python.org/}), Jupyter notebooks \citep{Jupyter:2016}, Xtreme Gradient Boosting \citep[XGB,][]{Chen:2016}, scikit-learn \citep{scikit-learn}, Keras \citep{Chollet:2015} and Tensorflow \citep{Tensorflow:2015}.

\end{acknowledgements}

\bibliographystyle{aa}
\bibliography{mybib}{}

\end{document}